\documentclass[iop]{emulateapj-rtx4}

\usepackage{graphicx}  
\usepackage{dcolumn}   
\usepackage{bm}        
\usepackage{amsfonts,amsmath,amssymb,mathrsfs}
\usepackage{color}

\usepackage{natbib,times}
\definecolor{orange}{rgb}{1,0.5,0}

\newcommand{\ie}{\textit{i.e.}~}
\newcommand{\eg}{\textit{e.g.}~}
\newcommand{\ms}{{\rm ms}}

\shorttitle{The missing link: Merging neutron stars can produce jets}
\shortauthors{Rezzolla et al.}

\begin{document}

\title{The missing link: Merging neutron stars naturally produce 
jet-like structures and can power short Gamma-Ray Bursts}

\author{Luciano {Rezzolla},\altaffilmark{1} Bruno
  {Giacomazzo},\altaffilmark{2,3} Luca {Baiotti},\altaffilmark{4}
  Jonathan {Granot},\altaffilmark{5} Chryssa
  {Kouveliotou},\altaffilmark{6} Miguel A. {Aloy},\altaffilmark{7}}

\altaffiltext{1}{Max-Planck-Institut f\"ur Gravitationsphysik, Albert-Einstein-Institut, Potsdam, Germany}

\altaffiltext{2}{Department of Astronomy, University of Maryland, College Park, MD, USA}
\altaffiltext{3}{Gravitational Astrophysics Laboratory, NASA Goddard Space Flight Center, Greenbelt, MD, USA}

\altaffiltext{4}{Institute of Laser Engineering, Osaka University, Suita, Japan}

\altaffiltext{5}{Centre for Astrophysics Research, University of
    Hertfordshire, College Lane, Hatfield, Herts, AL10 9AB, UK}

\altaffiltext{6}{Space Science Office, VP62, NASA/Marshall Space Flight Center, Huntsville, AL 35812, USA}

\altaffiltext{7}{Departamento de Astronom\'ia y Astrofis\'ica, Universidad de Valencia, 46100-Burjassot (Valencia), Spain}

\begin{abstract}
  Short gamma-ray Bursts (SGRBs) are among the most luminous
  explosions in the universe, releasing in less than one second the
  energy emitted by our Galaxy over one year. Despite decades of
  observations, the nature of their ``central engine'' remains
  unknown. Considering a binary of magnetized neutron stars
  and solving the Einstein equations, we show that their merger results in
  a rapidly spinning black hole surrounded by a hot and highly
  magnetized torus. Lasting over $35\;$ms and much longer than
  previous simulations, our study reveals that magnetohydrodynamical
  instabilities amplify an initially turbulent magnetic field of
  $\sim10^{12}\;$G to produce an ordered poloidal field of
  $\sim10^{15}\;$G along the black-hole spin-axis, within a
  half-opening angle of $\sim30^\circ$, which may naturally launch a
  relativistic jet. The broad consistency of our \textit{ab-initio}
  calculations with SGRB observations shows that the merger of
  magnetized neutron stars can provide the basic physical conditions
  for the central engine of SGRBs.
\end{abstract}


\maketitle

\section{Introduction}

The numerical investigation of the inspiral and merger of binary
neutron stars (NSs) in full general relativity has made big strides in
recent years. Crucial improvements in the formulation of the equations
and numerical methods, along with increased computational resources,
have extended the scope of early simulations. These developments have
made it possible to compute the full evolution, from large
binary-separations up to black-hole (BH) formation, without and with
magnetic fields~\citep{Shibata06, Baiotti08, Anderson2008, Liu2008,
  Giacomazzo2009, Giacomazzo2011}, and with idealized or realistic
equations-of-state \citep[EOS;][]{Rezzolla:2010,Kiuchi2010}. This
tremendous progress is also providing information about the entire
gravitational waveform, from the early inspiral up to the ringing of
the BH (see, \eg~\citet{Duez2010,Baiotti10}). Advanced interferometric
detectors starting from 2014 are expected to observe these sources at
a rate of $\sim40-400$ events per year~\citep{Abadie:2010}.

These simulations also probe whether the end-product of mergers can
serve as the ``central engine'' of
SGRBs~\citep{Paczynski86,Eichler89,Narayan92}. The prevalent
scenario invoked to explain SGRBs involves the coalescence of a binary
system of compact objects, \eg a BH and a NS or two
NSs~\citep{RuffertJanka1999,Rosswog:2003,Nakar:2007yr,Lee:2007js}. After
the coalescence, the merged object is expected to collapse to a BH
surrounded by an accretion torus. An essential ingredient in this
scenario is the formation of a central engine, which is required to
launch a relativistic outflow with an energy of
$\sim10^{48}-10^{50}$~erg on a timescale of
$\sim0.1-1$~s~\citep{Nakar:2007yr,Lee:2007js}. With only one possible
exception~\citep{DePasquale:2010}, SGRB afterglows do not clearly show
a jet-associated light-curve steepening~\citep{Nakar:2007yr}, thus
suggesting typical jet opening half-angles of at least several
degrees.

The qualitative scenario described above is generally supported by the
association of SGRBs with old stellar populations, distinct from the
young massive star associations for long
GRBs~\citep{Fox2005,Prochaska06}. It is also supported to a good
extent by fully general-relativistic simulations, which show that the
formation of a torus of mass $M_{\text{tor}}\lesssim0.4\,M_{\odot}$
around a BH with spin $J/M^2\simeq0.7-0.8$, is
inevitable~\citep{Rezzolla:2010}. However, the simulations have so far
failed to show that a jet \textit{can} be produced. We here provide
the first evidence that the merger of a binary of modestly magnetized
NSs naturally forms many of the conditions needed to produce a jet of
ultrastrong magnetic field, with properties that are broadly
consistent with SGRB observations.

\section{Numerical Simulations}

\begin{figure*}
  \begin{center}
     \includegraphics[angle=0,width=7.0cm]{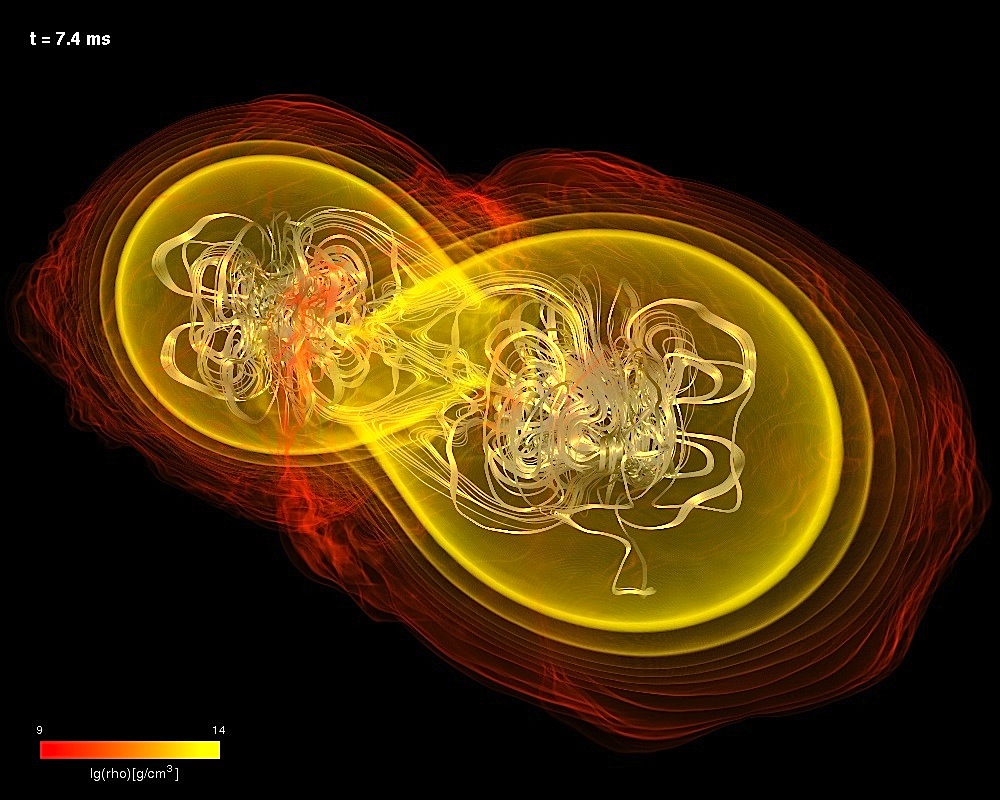}
     \hskip 0.2cm
     \includegraphics[angle=0,width=7.0cm]{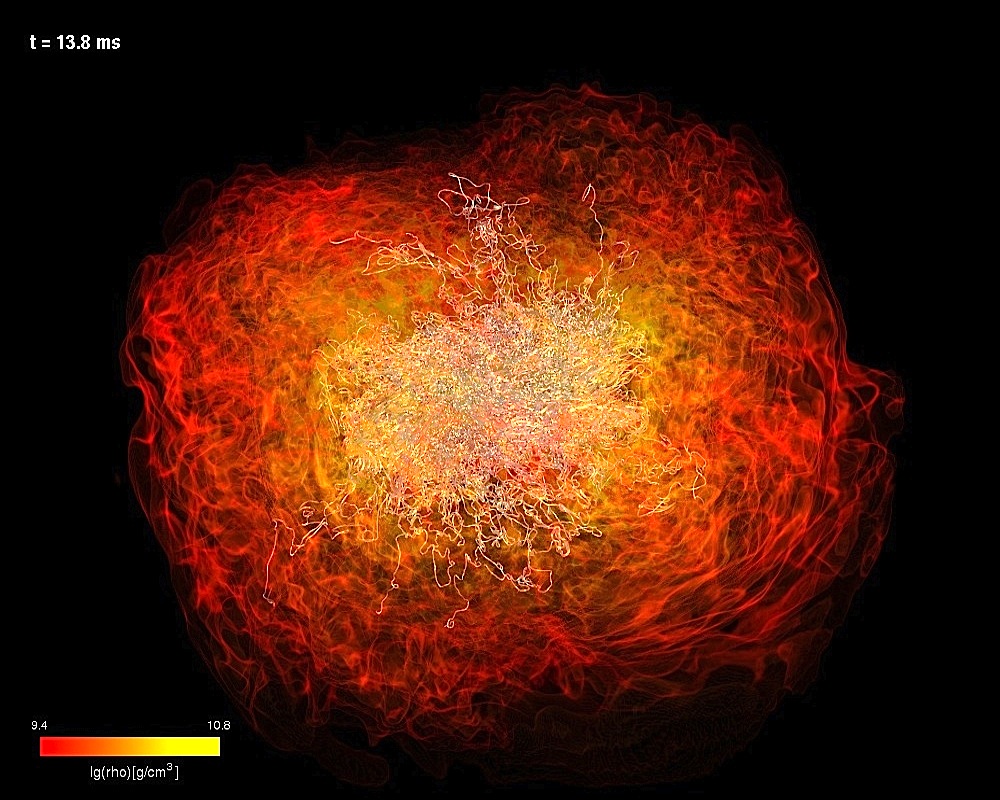}
     \vskip 0.2cm
     \includegraphics[angle=0,width=7.0cm]{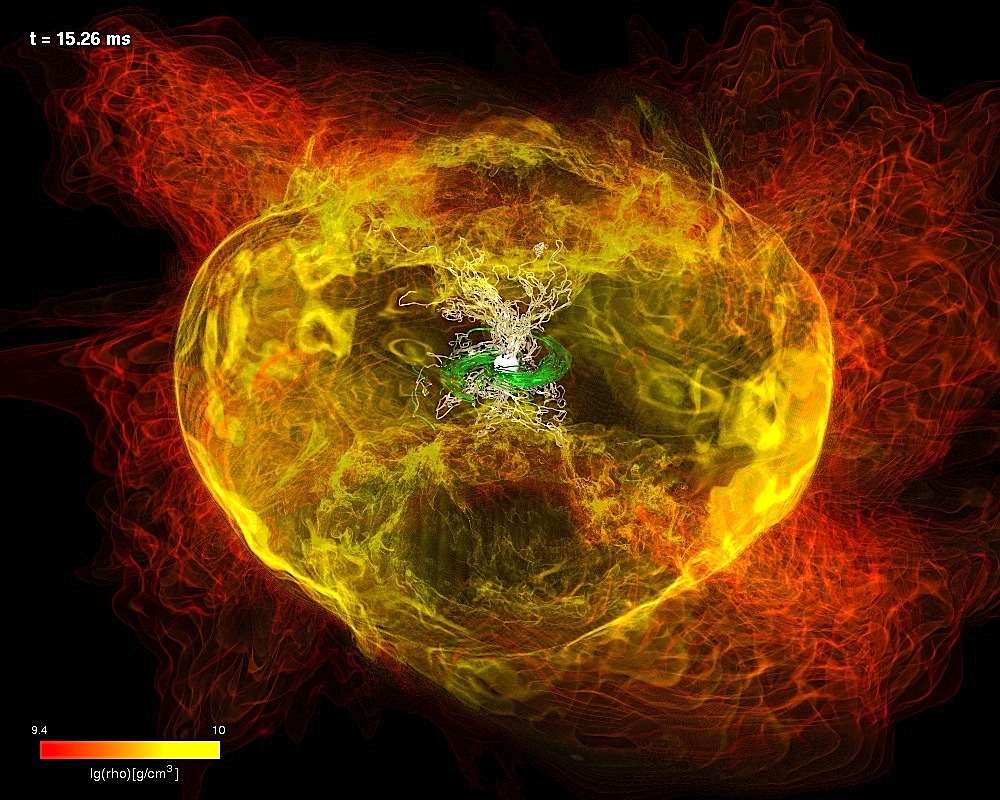}
     \hskip 0.2cm
     \includegraphics[angle=0,width=7.0cm]{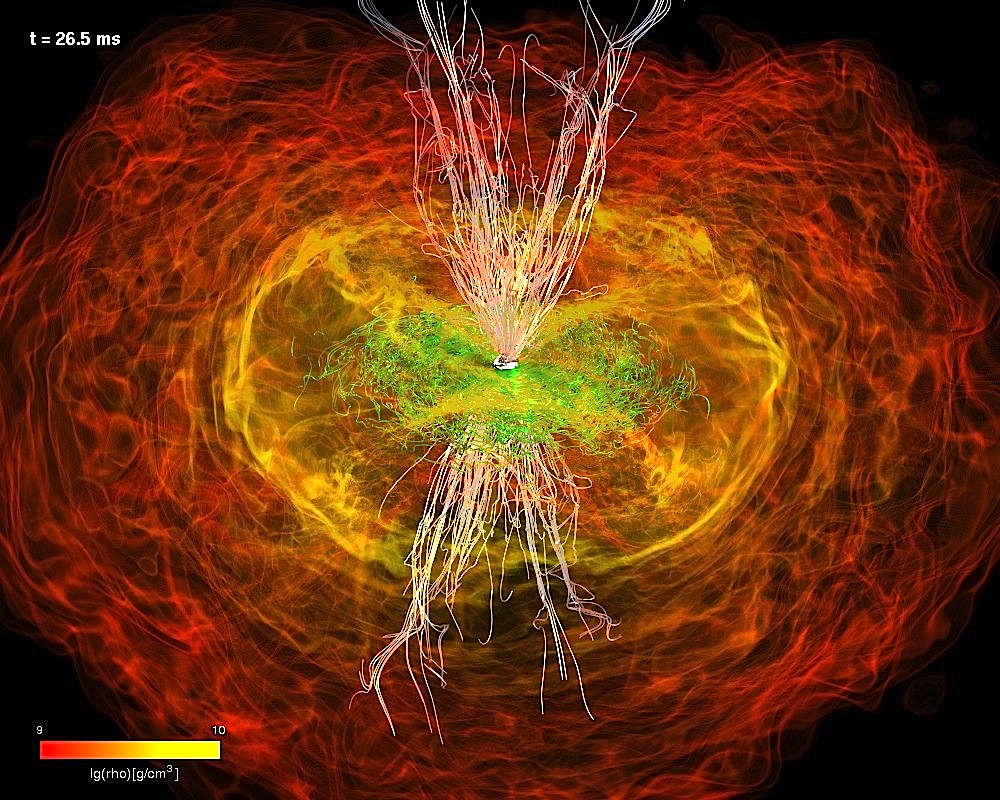}
  \end{center} \caption{Snapshots at representative times of the
  evolution of the binary and of the formation of a large-scale
  ordered magnetic field. Shown with a color-code map is the density,
  over which the magnetic-field lines are superposed. The panels in
  the upper row refer to the binary during the merger ($t=7.4\,\ms$)
  and \textit{before} the collapse to BH ($t=13.8\,\ms$), while those
  in the lower row to the evolution \textit{after} the formation of
  the BH ($t=15.26\,\ms$, $t=26.5\,\ms$). Green lines sample the
  magnetic field in the torus and on the equatorial plane, while white
  lines show the magnetic field outside the torus and near the BH spin
  axis. The inner/outer part of the torus has a size of
  $\sim90/170\,{\rm km}$, while the horizon has a diameter of
  $\simeq9\,{\rm km}$.} \label{fig:fig1}
\end{figure*}

For the simulations we use the Cactus/Carpet/Whisky codes
\citep[]{Schentter2004, Thornburg2004, Giacomazzo2007, Pollney07} 
and we consider a configuration that could represent the properties of a
NS-binary a few orbits before its coalescence, within a fully
general-relativistic and an ideal-magnetohydrodynamic (MHD)
framework~\citep{Giacomazzo2007,Giacomazzo2011}. More specifically, we
simulate two equal-mass NSs, each with a gravitational mass of
$1.5\,M_{\odot}$ (\ie sufficiently large to produce a BH soon after
the merger), an equatorial radius of $13.6\,{\rm km}$, and on a
circular orbit with initial separation of $\simeq45\,{\rm km}$ between
the centres (all lengthscales are coordinate scales;~\citep{Taniguchi02}). Confined in each
star is a poloidal magnetic field with a maximum strength of
$10^{12}\,{\rm G}$~\citep[indicated as \texttt{M1.62-B12}
  in][]{Giacomazzo2011}. At this separation, the binary loses energy
and angular momentum via emission of gravitational waves (GWs), thus
rapidly proceeding on tighter orbits as it evolves. After about
$8\,\ms$ ($\sim3$ orbits) the two NSs merge forming a hypermassive NS
(HMNS), namely, a rapidly and differentially-rotating NS, whose mass,
$3.0\,M_{\odot}$, is above the maximum mass, $2.1\,M_{\odot}$, allowed
with uniform rotation by our ideal-gas EOS\footnote{The use of a
  simplified EOS does not influence particularly our results
  besides determining the precise time when the HMNS collapses to a
  BH.}  with an adiabatic index of 2. Being metastable, a HMNS can
exist as long as it is able to resist against collapse via a suitable
redistribution of angular momentum \citep[\eg deforming into a ``bar''
  shape,][]{Shibata06,Baiotti08}, or through the pressure support 
coming from the large temperature-increase produced
by the merger. However, because the HMNS is also losing angular
momentum through GWs, its lifetime is limited to a few ms, after which
it collapses to a BH with mass $M=2.91\,M_{\odot}$ and spin
$J/M^2=0.81$, surrounded by a hot and dense torus with mass
$M_{\text{tor}}=0.063\,M_{\odot}$~\citep{Giacomazzo2011}.

\section{Dynamics of matter and magnetic fields}

These stages of the evolution can be seen in Figure~\ref{fig:fig1},
which shows snapshots of the density color-coded between $10^9$ and
$10^{10}\,{\rm gr/cm}^3$, and of the magnetic field lines (green on
the equatorial plane and white outside the torus). Soon after the BH
formation the torus reaches a quasi-stationary regime, during which
the density has maximum values of $\sim10^{11}\,{\rm g/cm}^3$, while
the accretion rate settles to $\dot{M}\simeq0.2\,M_{\odot}/{\rm
  s}$. Using the measured values of the torus mass and of the
accretion rate, and assuming the latter will not change significantly,
such a regime could last for $t_{\text{accr}}=
M_{\text{tor}}/\dot{M}\simeq0.3$~s, after which the torus is fully
accreted; furthermore, if the two NSs have unequal masses, tidal tails
are produced which provide additional late-time
accretion~\citep{Rezzolla:2010}. This accretion timescale is close to
the typical observed SGRB
durations~\citep{Kouveliotou1993,Nakar:2007yr}.  It is also long
enough for the neutrinos produced in the torus to escape and
annihilate in its neighborhood; estimates of the associated energy
deposition rate range from $\sim10^{48}\,{\rm
  erg/s}$~\citep{Dessart2009} to $\sim10^{50}\,{\rm
  erg/s}$~\citep{Setiawan2004}, thus leading to a total energy
deposition between a few $10^{47}\,{\rm erg}$ and a few $10^{49}\,{\rm
  erg}$ over a fraction of a second. This energy would be sufficient
to launch a relativistic fireball, but because we do not yet account
for radiative losses, the large reservoir of thermal energy in the
torus cannot be extracted in our simulations.

\begin{figure*}
  \begin{center}
     \includegraphics[angle=0,width=7.0cm]{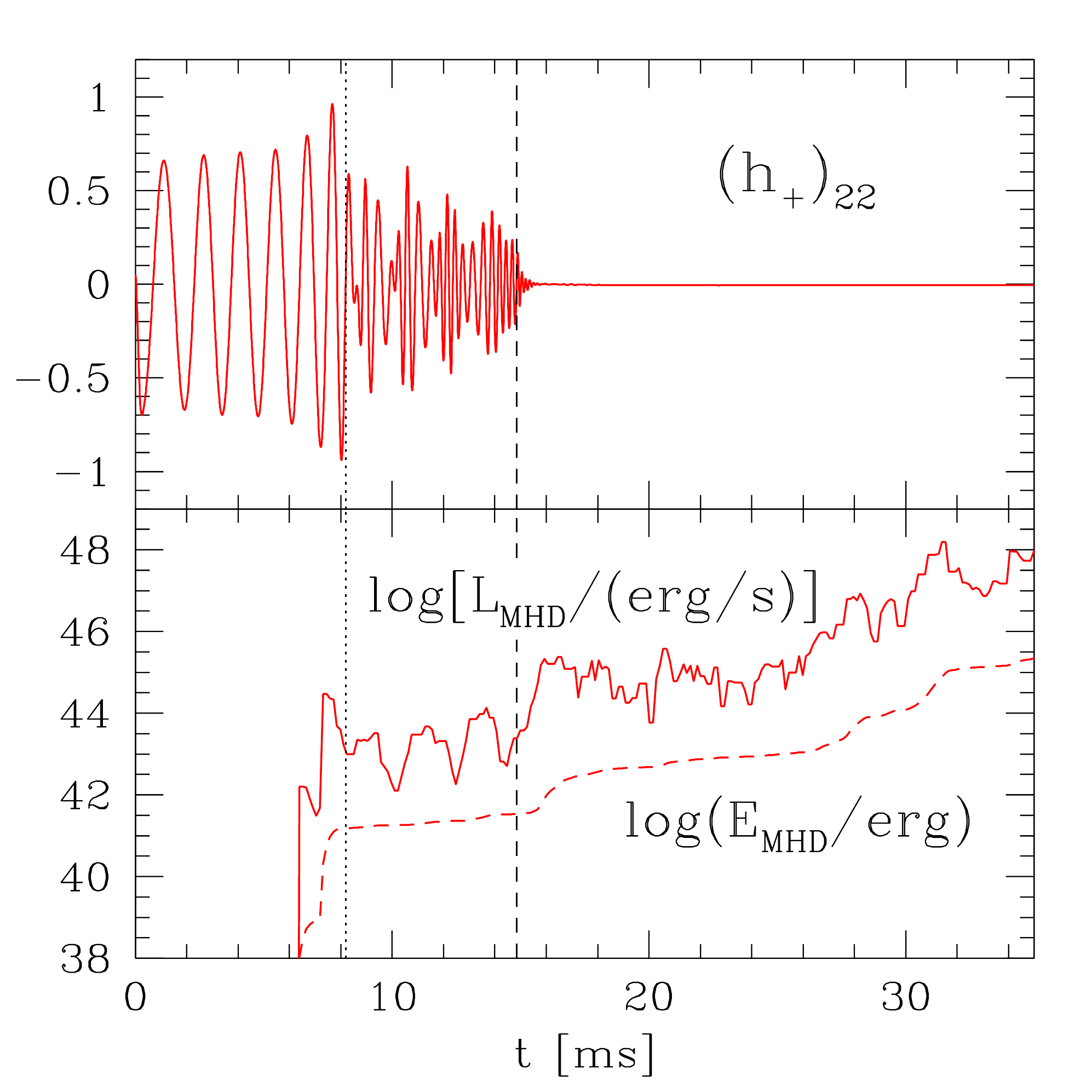}
     \hskip 1.0cm
     \includegraphics[angle=0,width=7.0cm]{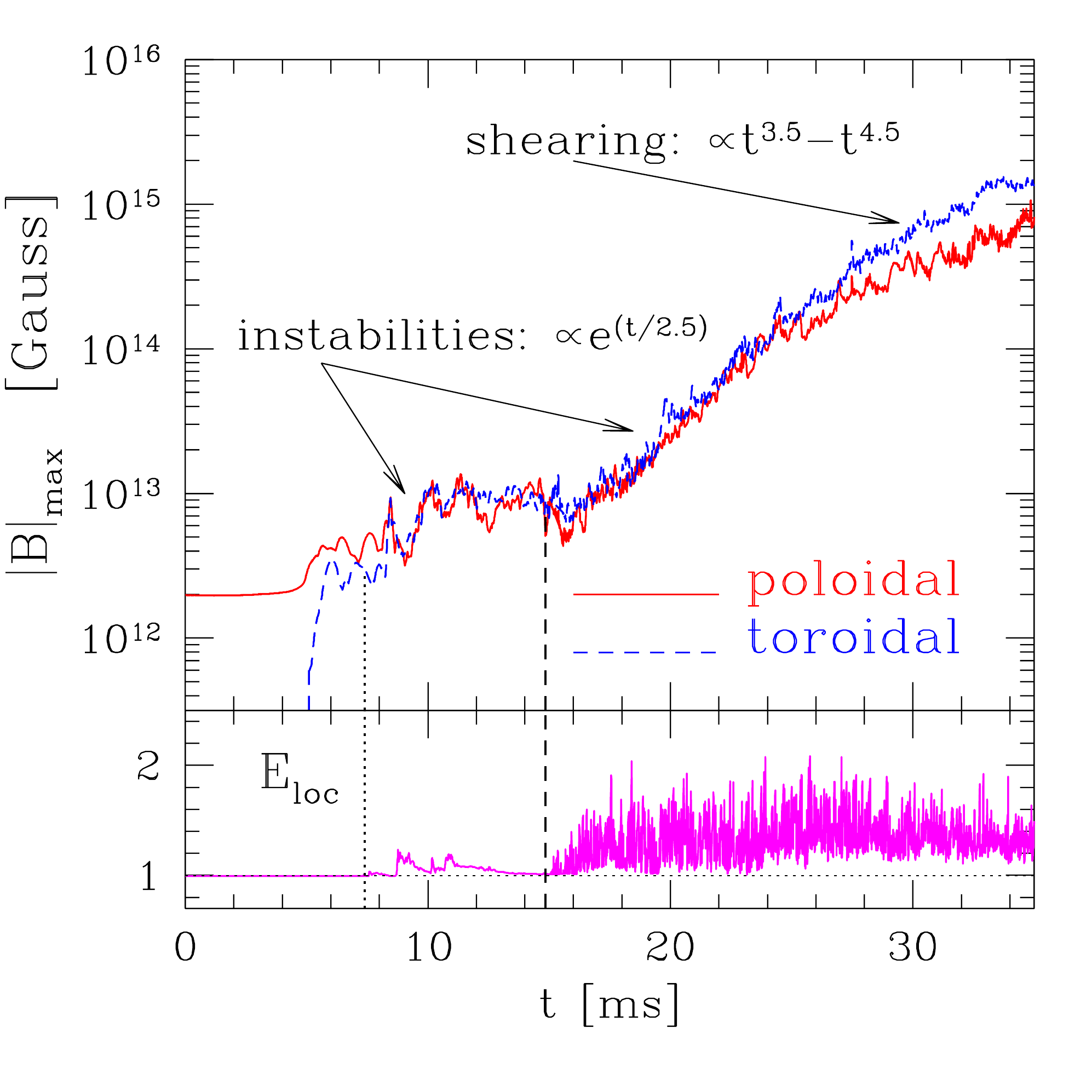}
  \end{center}
  \caption{\textit{Left panel:} GW signal shown through the
    $\ell=2,m=2$ mode of the $+$ polarization, $(h_+)_{22}$, (top
    part) and the MHD luminosity, $L_{\rm MHD}$, (bottom part) as
    computed from the integrated Poynting flux and shown with a solid
    line. The corresponding energy, $E_{\rm MHD}$, is shown with a
    dashed line. The dotted and dashed vertical lines show the times
    of merger (as deduced from the first peak in the evolution of
    amplitude of the GW amplitude) and BH formation,
    respectively. \textit{Right panel:} Evolution of the maximum of
    the magnetic field in its poloidal (red solid line) and toroidal
    (blue dashed line) components. The bottom panel shows the maximum
    local fluid energy indicating that an unbound outflow (\ie
    $E_{\text{loc}} > 1$) develops and is sustained after BH
    formation.}
  \label{fig:fig2}
\end{figure*}

The GW signal of the whole process is shown in the left panel of
Figure~\ref{fig:fig2}, while the bottom part exhibits the evolution of
the MHD luminosity, $L_{\rm MHD}$, as computed from the integrated
Poynting flux (solid line) and of the corresponding energy, $E_{\rm
  MHD}$ (dashed line). Clearly, the MHD emission starts only at the
time of merger and increases exponentially after BH formation, when
the GW signal essentially shuts off. Assuming that the
quasi-stationary MHD luminosity is $\simeq 4\times10^{48}\,{\rm erg/s}$,
the total MHD energy released during the lifetime of the torus is
$\simeq 1.2\times10^{48}\,{\rm erg}$, which, if spread over an opening
half-angle of $\sim30^\circ$ (see discussion below), suggests a lower
limit to the isotropic equivalent energy in the outflow of
$\simeq 9\times10^{48}\,{\rm erg}$. While this is at the low end of the
observed distribution of gamma-ray energies for SGRBs, larger MHD
luminosities are expected either through the additional growth of the
magnetic field via the ongoing winding of the field lines in the disk
(the simulation covers only one tenth of $t_{\rm accr}$), or when
magnetic reconnection (which cannot take place within our ideal-MHD
approach), is also accounted for \citep[which may also increase the
gamma-ray efficiency; see, \eg][]{MU2010}.

The last two panels of Figure~\ref{fig:fig1} offer views of the
accreting torus after the BH formation. Although the \textit{matter}
dynamics is quasi-stationary, the last two panels clearly show that
the \textit{magnetic field} is not and instead evolves significantly.
It is only when the system is followed well after the formation of a
BH, that MHD instabilities develop and generate the central,
low-density, poloidal-field funnel. This regime, which was not
accessible to previous simulations~\citep{Price06, Anderson2008,
  Liu2008}, is essential for the jet formation~\citep{Aloy:2005,
  Komissarov:2009}. Because the strongly magnetized matter in the
torus is highly conductive, it shears the magnetic-field lines via
differential rotation. A measurement of the angular velocity in the
torus indicates that it is essentially Keplerian and thus unstable to
the magneto-rotational instability \citep[MRI;][]{BalbusHawley1998},
which develops $\simeq5\,\ms$ after BH formation and amplifies
exponentially both the poloidal and the toroidal magnetic fields; the
{\it e}-folding time of the instability is $\simeq2.5\,\ms$ and in good
agreement with the one expected in the outer parts of the
torus~\citep{BalbusHawley1998}. Because of this exponential growth,
the final value of the magnetic field is largely insensitive to the
initial strength and thus a robust feature of the dynamics.

A quantitative view of the magnetic-field growth is shown in the right
panel of Figure~\ref{fig:fig2}, which shows the evolution of the maximum
values in the poloidal and toroidal components. Note that the latter
is negligibly small before the merger, reaches equipartition with the
poloidal field as a result of a Kelvin-Helmholtz instability triggered
by the shearing of the stellar surfaces at
merger~\citep{Price06,Giacomazzo2009}, and finally grows to
$\simeq10^{15}$ G by the end of the simulation. At later times
($t\gtrsim22\,\ms$), when the instability is suppressed, the further
growth of the field is due to the shearing of the field lines and it
increases only as a power law with exponent $3.5~(4.5)$ for the
poloidal~(toroidal) component. Although the magnetic-field growth
essentially stalls after $t\simeq35\,\ms$, further slower growths are
possible~\citep{Obergaulinger:2009}, yielding correspondingly larger
Poynting fluxes. Indeed, when the ratio between the magnetic flux
across the horizon and the mass accretion rate becomes sufficiently
large, a Blandford-Znajek mechanism~\citep{Blandford1977} may be
ignited~\citep{Komissarov:2009dn}; such conditions are not met over
the timescale of our simulations, but could develop over longer
timescales. Also shown in the right panel of Figure~\ref{fig:fig2} is
the maximum local fluid energy, highlighting that an \textit{unbound
  outflow} (\ie $E_{\text{loc}}>1$) develops after BH formation along
the outer walls of the torus and persists for the whole duration of
the simulation.

\begin{figure*}
  \begin{center}
     \includegraphics[angle=0,width=7.0cm]{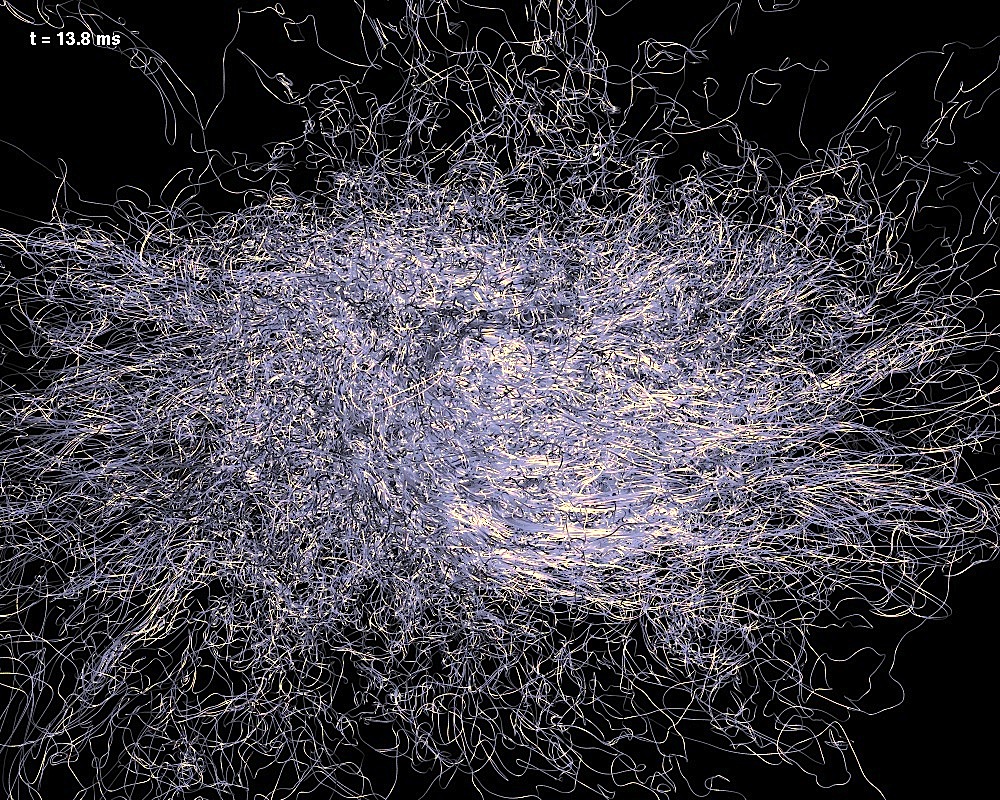}
     \hskip 0.2cm
     \includegraphics[angle=0,width=7.0cm]{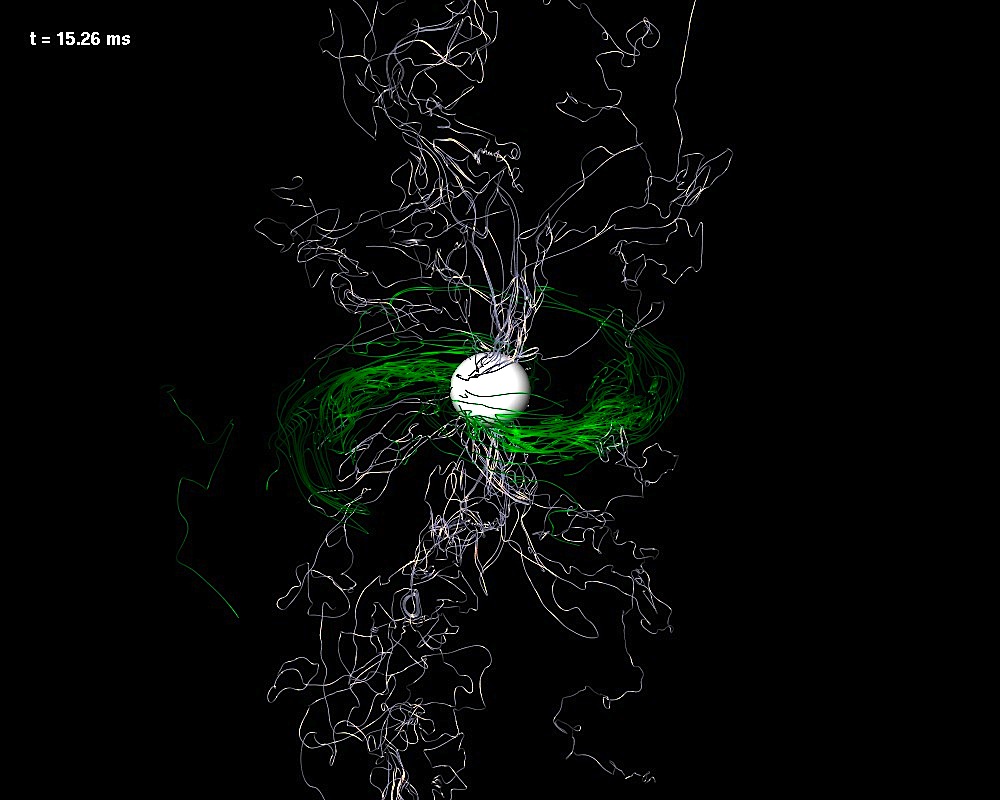}
     \vskip 0.2cm
     \includegraphics[angle=0,width=7.0cm]{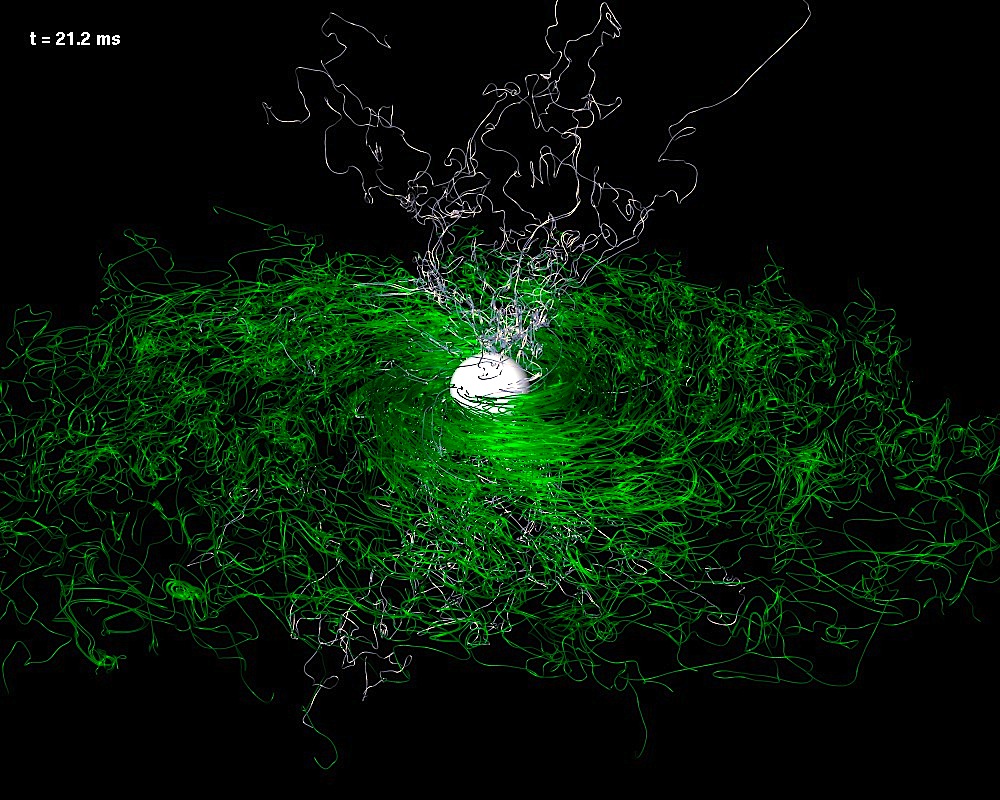}
     \hskip 0.2cm
     \includegraphics[angle=0,width=7.0cm]{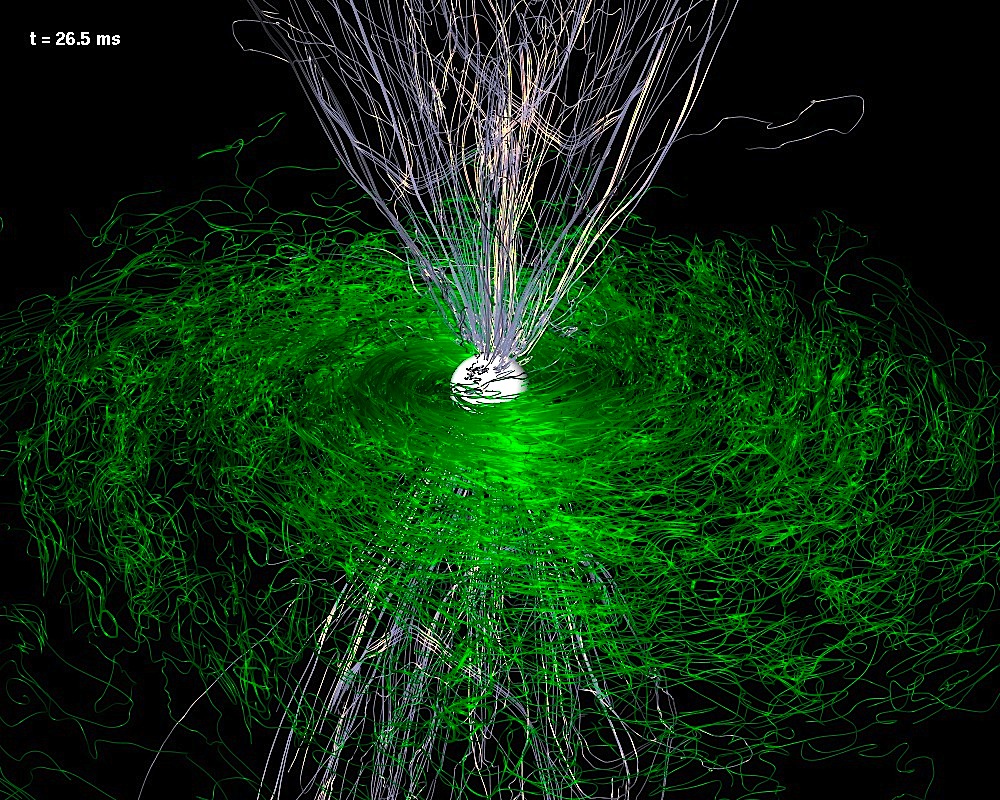}
  \end{center}
  \caption{Magnetic-field structure in the HMNS (first panel) and
    after the collapse to BH (last three panels). Green refers to
    magnetic-field lines inside the torus and on the equatorial plane,
    while white refers to magnetic-field lines outside the torus and
    near the axis. The highly turbulent, predominantly poloidal
    magnetic-field structure in the HMNS ($t=13.8\,\ms$) changes
    systematically as the BH is produced ($t=15.26\,\ms$), leading to
    the formation of a predominantly toroidal magnetic field in the
    torus ($t=21.2\,\ms$). All panels have the same linear scale, with
    the horizon diameter being of $\simeq9\,{\rm km}$.}
  \label{fig:fig4}
\end{figure*}

Finally, Figure~\ref{fig:fig4} provides a summary of the magnetic-field
dynamics. It shows the magnetic field in the HMNS formed after the
merger and its structure and dynamics after the collapse to BH. In
particular, in the last three panels it shows the magnetic-field
structure inside the torus and on the equatorial plane (green), and
outside the torus and near the axis (white). It is apparent that the
highly turbulent magnetic field in the HMNS ($t=13.8\,\ms$) changes
systematically as the BH is produced ($t=15.26\,\ms$), leading to the
formation of a toroidal magnetic field in the torus
($t=21.2\,\ms$). As the MRI sets in, the magnetic field is not only
amplified, but also organizes itself into a dual structure, which is
mostly toroidal in the accretion torus with $B_{\rm
  tor}\simeq 2\times10^{15}\,{\rm G}$, but predominantly poloidal and
jet-like along the BH spin axis, with $B_{\rm
  pol}\simeq 8\times10^{14}\,{\rm G}$ ($t=26.5\,\ms$). Note that the
generation of an ordered large-scale field is far from trivial and a
nonlinear dynamo may explain why the MRI brings a magnetic field
self-organization, as it has been also suggested in case of
MRI-mediated growth of the magnetic field in the conditions met in the
collapse of massive stellar cores~\citep{Lesur:2008,
  Obergaulinger:2009}. However, the jet-like structure produced in the
simulation is not yet the highly collimated ultrarelativistic outflow
expected in SGRBs (see also below).

The hollow jet-like magnetic structure has an opening half-angle of
$\sim30^\circ$, which sets an upper limit for the opening half-angle
of any potential outflow, either produced by neutrino energy
deposition~\citep{Aloy:2005} or by electromagnetic (EM)
processes~\citep{Komissarov:2009}.  In our simulations most of the
outflow develops along the edges of the jet-like structure, via a
turbulent layer of EM driven matter, which shields the central funnel
from excessive baryonic pollution. We envision that such a layer is
crucial to set the opening angle of any ultrarelativistic jet, to
shape both the radial and transverse structure of the jet, as well as
to determine its stability properties. The Lorentz factors of the
outflow measured in our simulations are not very high
($\Gamma\lesssim4$), but can potentially be amplified by several
orders of magnitude in the inner baryon-poor regions through
special-relativistic effects~\citep{Aloy:2006rd} or the variability of
the flow~\citep{Granot2011}. We expect that such accelerations will be
produced as a more realistic and general-relativistic treatment of the
radiative losses will become computationally affordable.

\section{Comparison with observations}

Below we briefly discuss how our results broadly match the properties
of the central engine as deduced from the observations.

\smallskip
\subsection{\bf Duration}
The observed duration of the prompt
gamma-ray emission GRBs is energy dependent and is usually determined
through $T_x$, the time over which $x\%$ of the total counts are
observed, between the $(100-x)/2$ and $(100+x)/2$ percentiles. The
most common intervals used are $T_{90}$ (or $T_{50}$), initially
defined~\citep{Kouveliotou1993} between $20\;$keV and $2\;$MeV. The
GRB duration distribution is bimodal~\citep{Kouveliotou1993}, where
the durations of SGRBs (approximately 25\% of GRBs) are well-fit by a
fairly wide log-normal distribution centered around $T_{90}\approx
0.8\;$s with a FWHM of 1.4 dex~\citep{Nakar:2007yr}. The typical
redshifts of the SGRBs observed with {\it Swift} are in the range $z
\sim0.3-1$, suggesting a central value of the intrinsic duration
distribution of $\approx (1+z)^{-1}0.8\;{\rm
  s}\sim0.5\;$s, and a comparably wide distribution around this
value. This is in close agreement with our accretion time of
$\sim0.3\;$s.

\smallskip
\subsection{\bf Energy} The isotropic equivalent energy output
in the prompt gamma-ray emission of SGRBs, $E_{\rm\gamma,iso}$, spans
a wide range, from $(2.7\pm1)\times 10^{48}\;$erg (in the observed
energy range $15-350\;$keV) for GRB~050509B at a redshift of $z =
0.225$~\citep{GRB050509B}, up to $(1.08\pm0.06)\times 10^{53}\;$erg
(in the observed energy range 10$\;$keV -- 30$\;$GeV) for GRB~090510
at $z=0.903$~\citep{GRB090510}. However, the most typical values are
in the range
$E_{\rm\gamma,iso}\sim10^{49}-10^{51}\;$erg~\citep{Nakar:2007yr}.  In
our model, the highly relativistic outflow may be powered either by
neutrino-anti neutrino annihilation, or by the Blandford-Znajek
mechanism. For the former one might expect a total energy release
between a few $10^{47}\;$erg and
$\sim10^{49}\;$erg~\citep{Oechslin:2006,Birkl:2007}, into a bipolar
relativistic jet of opening half-angle $\theta_{\rm
  jet}\sim8-30^\circ$, corresponding to a fraction $f_b\sim0.01-0.13$
of the total solid angle, and isotropic equivalent energies, $E_{\rm
  \nu\bar{\nu},iso}$, between a few $10^{48}\;$erg and
$\sim10^{51}\;$erg. For the latter mechanism, instead, and if the
magnetization near the event horizon becomes sufficiently high, we
could expect a jet power for our values for the BH mass and spin that
is of~\citep{Lee:2000,Perez:2010}
\begin{equation}
L_{\rm BZ} \simeq 3.0\times10^{50}\left(\frac{f_{\rm rel}}{0.1}\right)
\left(\frac{B}{2\times10^{15}\,\rm{G}}\right)^2\, {\rm erg/s}\,,
\end{equation}
where $f_{\rm rel}$ is the fraction of the total Blandford-Znajek
power that is channeled into the resulting relativistic jet (and
$f_{\rm rel}\sim0.1$ might be expected for ejecta with asymptotic
Lorentz factors above $100$). This relativistic outflow is launched
over a timescale of $\sim0.2\,{\rm s}$ and corresponds to
\begin{equation}
E_{\rm BZ, iso}\simeq 1.2\times10^{51} \left(\frac{f_{\rm rel}}{0.1}\right)
\left(\frac{f_b}{0.05}\right)^{-1}
\left(\frac{B}{2\times 10^{15}\,\rm{G}}\right)^2\,{\rm erg}\,.
\end{equation}

Comparing the X-ray afterglow luminosity (after 10 or 11 hours) and
$E_{\rm\gamma,iso}$ suggests that the efficiency of the prompt
gamma-ray emission in SGRBs is typically high
~\citep{GRB050509B,GR-RF09}, and similar to that of long GRBs
~\citep{GKP06}, with $E_{\rm\gamma,iso}\sim(0.1-0.9)E_{\rm
  iso}$, radiating between $\sim10\%$ and $\sim90\%$ of the initial
energy of the ultrarelativistic outflow. Therefore, our model is able
to accommodate the observed $E_{\rm\gamma,iso}$ values.

\smallskip
\subsection{\bf Lorentz factor} The Fermi Gamma-Ray Space
Telescope has detected GeV emission from
SGRBs~\citep{GRB080825C}, suggesting typical lower limits
of $\Gamma_{\rm min}\sim10^2-10^3$. In particular, $\Gamma_{\rm
  min}\approx1200$ was obtained for
GRB~090510~\citep{GRB090510}. However, a more realistic model
~\citep{Granot08} results in $\Gamma_{\rm min}$ values lower by a
factor of $\sim3$ \citep{GRB090926a}. Therefore, the central engine
should be capable of producing outflow Lorentz factors of at least a
few hundred. The fact that our simulation produces a strongly
magnetized mildly relativistic outflow at angles near $\sim30^\circ$
from the BH spin axis would help shield the inner region near the spin
axis from excessive baryon loading, and thus assist in achieving high
asymptotic Lorentz factors at large distance from the source, after
the outflow in this region is triggered by neutrinos and/or the
Blandford-Znajek mechanism.

\smallskip
\subsection{\bf Jet angular structure} This is poorly
constrained by observations (even more so than for long GRBs). The
only compelling case for a jet break in the afterglow light curve is
for GRB~090510~\citep{DePasquale:2010}, which occurred very early on
(after $\sim1400\;$s), and would thus imply an extremely narrow jet
($\theta_{\rm jet}\sim0.2-0.4^\circ$) and modest true energy output
in gamma rays ($\sim10^{48}\;$erg). If this is indeed a jet break, it
might correspond to a line of sight near a very narrow and bright core
of a jet, which also has significantly wider wings. Observers with
lines of sight along these wings would then see a much dimmer and more
typical SGRB~\citep{Rossi02,Peng05,Racusin08}; without such wings,
however, the observations would suggest a very large intrinsic and
beaming-corrected event rate per unit volume. In most cases there are
only lower limits on a possible jet break time~\citep{Nakar:2007yr},
resulting in typical limits of $f_b \gtrsim 10^{-2}$ or $\theta_{\rm
  jet} \gtrsim 8^\circ$. This is consistent with our expectation of
$\theta_{\rm jet}\sim8-30^\circ$ for the ultrarelativistic ejecta
capable of producing a SGRB (which would also imply a reasonable SGRB
intrinsic event rate per unit volume).

\section{Conclusions}

The calculations reported here clearly demonstrate that a binary
merger of two NSs inevitably leads to the formation of a relativistic
jet-like and ultrastrong magnetic field, which could serve as a
central engine for SGRBs. Because the magnetic-field growth is
exponential, the picture emerging from our simulations is rather
general and applies equally even to mildly magnetized NSs. Overall,
this work removes a significant earlier uncertainty as to whether such
binary mergers can indeed produce the central engines of SGRBs. While
the EM energy release is already broadly compatible with the
observations, our simulations lack a proper treatment of the energy
losses via photons and neutrinos, which can provide a fundamental
contribution to the energy input necessary to launch the fireball and
cool the torus~\citep{Setiawan2004,Dessart2009}. This additional
energy input, whose self-consistent inclusion in general relativity
remains extremely challenging, may help to launch an ultrarelativistic
outflow very early after the BH forms and complete the picture of the
central engine of a SGRB.
 
\acknowledgements It is a pleasure to thank Michael Koppitz, Ian
Hinder and Neil Gehrels for useful input. Partial support comes from
the DFG grant SFB/Transregio~7, by ``CompStar'', a Research Networking
Programme of the European Science Foundation, by the MEXT Grant-in-Aid
for Young Scientists (22740163), by the NASA grant number NNX09AI75G,
and by the European Research Council grant number 259276-CAMAP.
J.G. gratefully acknowledges a Royal Society Wolfson Research Merit
Award.

\end{document}